\documentclass[10pt,a4paper,twoside]{article}
\usepackage{epsfig}
\usepackage{baltlat6}
\usepackage{array}
\usepackage{here}
\usepackage{natbib}
\usepackage[dvipdfm=true]{hyperref}

\makeatletter
\def\captionb#1#2{\refstepcounter{figure}\small\vspace{2mm}
\vbox{\baselineskip=9.5pt\hhuad{\bfseries Fig.\,#1.}
 \,#2} \baselineskip=11pt}

\def\captiontb#1#2{\refstepcounter{table}\small\vspace{2mm}
\vbox{\baselineskip=9.5pt\hhuad{\bfseries Table\,#1.}
 \,#2} \baselineskip=11pt}

\def\sectionb#1#2{\refstepcounter{section}\vspace{5mm}\hbox{\kern-.9pt
{#1.\ \ }\vtop{\noindent #2}\nopagebreak}
\vspace{1mm} \noindent\baselineskip=11pt}

\def\subsectionb#1#2{\refstepcounter{subsection}\vspace{3mm}\hbox{\kern-.9pt
{\it #1.\ \ }\vtop{\noindent\it #2}\nopagebreak}
\vspace{.5mm} \noindent\baselineskip=11pt}

\def\subsubsectionb#1#2{\refstepcounter{subsubsection}\vspace{3mm}\hbox{\kern-.9pt
{#1.\ \ }\vtop{\noindent #2}\nopagebreak}
\vspace{.5mm} \noindent\baselineskip=11pt}
\makeatother

\begin{document}
\ \
\vspace{0.5mm}
\setcounter{page}{151}
\vspace{8mm}

\titlehead{Baltic Astronomy, vol.\,23, 151--170, 2014}

\titleb{HOW DOES THE MASS TRANSPORT IN DISK GALAXY \\ MODELS INFLUENCE
THE CHARACTER OF ORBITS?}

\begin{authorl}
\authorb{Euaggelos E. Zotos}{}
\end{authorl}

\moveright-3.2mm
\vbox{
\begin{addressl}
\addressb{}{Department of Physics, School of Science, Aristotle University of Thessaloniki, \\
GR-541 24, Thessaloniki, Greece; \ e-mail: evzotos@physics.auth.gr}
\end{addressl}}

\submitb{Received: 2014 August 29; accepted: 2014 September 18}

\begin{summary} We explore the regular or chaotic nature of orbits of
stars moving in the meridional $(R,z)$ plane of an axially symmetric
time-dependent disk galaxy model with a central, spherically symmetric
nucleus.  In particular, mass is linearly transported from the disk to
the galactic nucleus, in order to mimic, in a way, the case of
self-consistent interactions of an actual N-body simulation.  We thus
try to unveil the influence of this mass transportation on the different
families of orbits of stars by monitoring how the percentage of
chaotic orbits, as well as the percentages of orbits of the main regular
resonant families, evolve as the galaxy develops a dense and massive
nucleus in its core.  The SALI method is applied to samples of orbits in
order to distinguish safely between ordered and chaotic motion.  In
addition, a method based on the concept of spectral dynamics is used for
identifying the various families of regular orbits and also for
recognizing the secondary resonances that bifurcate from them.  Our
computations strongly suggest that the amount of the observed chaos is
substantially increased as the nucleus becomes more massive.
Furthermore, extensive numerical calculations indicate that there are
orbits which change their nature from regular to chaotic and vice versa
and also orbits which maintain their orbital character during the
galactic evolution.  The present outcomes are compared to earlier
related work.  \end{summary}


\resthead{The influence of mass transport in disk galaxy models}
{Euaggelos E. Zotos}

\sectionb{1}{INTRODUCTION}
\label{intro}

A convenient way for studying autonomous Hamiltonian systems is by
choosing some values of the total orbital energy, for which the position
as well as the extent of ordered and chaotic domains are
time-independent (TI) and can be accurately determined by a variety of
dynamical methods, especially in the case of low degrees of freedom.
The dynamical structure however, in such time-independent Hamiltonian
systems is usually very complex due to the presence of ``weak" and
``strong" chaos.  In addition, the motion of test particles takes place
on invariant tori of multiple dimensions and it can also display
surprising localization properties in both configuration and phase
space.  It is therefore natural to assume that time-dependent (TD)
Hamiltonian systems are expected to exhibit a much more complicated
nature since all the above-mentioned attributes evolve with time due to
the absence of any kind of integrals of motion (i.e., the total energy
integral) which are valid only in TI systems.  Indeed, it is true that
in TI systems orbits maintain their orbital character:  if for example,
an orbit is initially regular it will always remain so, while if it is
chaotic it might get trapped near the boundaries of stability regions
for relatively long time intervals (this phenomenon is known as ``sticky
motion") but it will certainty never relinquish its chaoticity.  On the
other hand, in TD Hamiltonians sudden transitions form regularity to
chaoticity and vice versa is a common behavior for individual
trajectories during their time-evolution.

The determination of ordered and chaotic properties of motion in
time-dependent galactic as well as cosmological models constitutes an
extended research area in the field on non-linear dynamics.  In a
previous work, \citet{CP03} used a simple analytic time-dependent model
in order to study the transition from order to chaos in a galaxy, when
mass is exponentially transported from the disk to the galactic core
thus forming a dense and massive spherical nucleus.  They found that
during the galactic evolution a large portion of low angular momentum
stars change their orbital nature from regular to chaotic.  The
investigation was continued in more detail in \citet{Z12} where it was
proved by conducting a systematic and thorough exploration of the phase
space that during the mass transportation stars do not change their
character only from regular to chaotic.  In fact, there is a
considerable amount of stars which maintain their orbital nature, while
there is also a small fraction of star orbits that change their
character from chaotic to regular.  Furthermore, \citet{PC06} revealed
the orbital behavior in a time-dependent double-barred galaxy model
when mass is transported from the primary bar to the inner disk
reporting that the galactic evolution significantly affects the nature
of orbits.  The interplay between ordered and chaotic motion in a
time-dependent Hamiltonian describing a barred galaxy was examined in
\citet{MBS13}, where it was observed that the percentage of chaos
increases when a linear mass transportation from the disk to the bar
takes place.  In the same vein, \citet{MM14} constructed an analytical
time-dependent potential, modeling the gravitational potentials of disk,
a bar and a dark matter halo, whose time-dependent parameters are
derived from a simulation.

Knowing whether the orbits are regular or chaotic is only the first step
toward the understanding of the overall behavior of the disk galaxy.
The second and beyond any doubt the most interesting step is the
distribution of regular orbits into different families.  In our study
therefore, once the orbits have been characterized as regular or
chaotic, we then further classified the regular orbits into different
families by using a frequency analysis method (\citealt{CA98,MCW05}).
Initially, \citet{BS82,BS84} proposed a technique, dubbed spectral
dynamics, for this particular purpose.  Later on, this method has been
extended and improved by \citet{SN96} and \citet{CA98}.  In the recent
work of \citet{ZC13} the algorithm was refined even further so it can be
used to classify orbits in the meridional plane.  In general terms, this
method computes the Fourier transform of the coordinates of an orbit,
identifies its peaks, extracts the corresponding frequencies, and
searches for the fundamental frequencies and their possible resonances.
Thus, we can easily identify the various families of regular orbits and
also recognize the secondary resonances that bifurcate from them.  This
technique has been successfully applied in several previous papers
(e.g., \citealt{ZC13,CZ13,ZCar13,ZCar14,Z14a,Z14b}) in the field of orbit
classification (not only regular versus chaotic, but also separating
regular orbits into different regular families) in different galactic
gravitational potentials.

The motivation of the present work is to apply the time-dependent
version of the disk galaxy model used in \citet{ZC13} in order to
examine the dynamical properties and the time-evolution of orbits as
mass is transferred from the disk to the central nucleus.  Moreover,
this time-dependent Hamiltonian can, in a way, mimic certain realistic
aspects that arise in N-body simulations.  Our paper is organized as
follows:  in Section \ref{galmod}, we explain in detail the properties
of the time-dependent disk galaxy model, while Section \ref{cometh} is
devoted to the description of the computational methods we used in order
to determine the character as well as the classification of the orbits.
A thorough numerical analysis regarding the influence of the mass
transport to the percentages of the different families of orbits is
performed in Section \ref{numres}.  Our paper ends with Section
\ref{conc}, where the main conclusions of our numerical investigation
are presented.

\sectionb{2}{DESCRIPTION OF THE GALACTIC MODEL}
\label{galmod}

Our aim is to determine the interplay between ordered and chaotic motion
of stars moving in the meridional plane of a time-dependent axially
symmetric disk galaxy with a central spherically symmetric nucleus when
the parameters of the model potential evolve in time.  For this
purpose, we use the usual cylindrical coordinates $(R, \phi, z)$, where
$z$ is the axis of symmetry.

The total gravitational potential $\Phi(R,z)$ consists of two components:
the nucleus potential $\Phi_{\rm n}$ and the flat disk potential
$\Phi_{\rm d}$. For the description of the spherically symmetric central
nucleus, we use a Plummer potential (e.g.,
\citealt{BT08})
\begin{equation}
\Phi_{\rm n}(R,z) = \frac{- G M_{\rm n}}{\sqrt{R^2 + z^2 + c_{\rm n}^2}}.
\label{Vn}
\end{equation}
Here $G$ is the gravitational constant, while $M_{\rm n}$ and $c_{\rm
n}$ are the mass and the scale length of the nucleus, respectively.
This potential has been successfully used in the past to model and,
therefore, interpret the effects of the central mass component in a
galaxy (see e.g., \citealt{HN90,HPN93,Z12,ZC13,Z14a}).  At this point,
it must be emphasized that we do not include any relativistic effects,
because the nucleus represents a bulge rather than a black hole or any
other compact object.

The galactic disk, on the other hand, is represented by the well-known
Miyamoto-Nagai potential \citet{MN75}
\begin{equation}
\Phi_{\rm d}(R,z) = \frac{- G M_{\rm d}}{\sqrt{R^2 + \left(\alpha + \sqrt{h^2 + z^2}\right)^2}},
\end{equation}
where, $M_{\rm d}$ is the mass  of the disk, $\alpha$ is the scale
length of the disk and $h$ corresponds to the disk's scale height.

We use a system of galactic units where the unit of length is 1 kpc, the
unit of velocity is 10 km\,s$^{-1}$ and $G = 1$.  Thus, the unit of mass
is $2.325 \times 10^7 {\rm M}_\odot$, that of time is $0.9778 \times
10^8$ yr, the unit of angular momentum (per unit mass) is 10 km kpc
s$^{-1}$, and the unit of energy (per unit mass) is 100 km$^2$s$^{-2}$.
In these units, the values of the involved parameters are:  $c_{\rm n} =
0.25$, $\alpha = 3$, $h = 0.175$, while $M_{\rm n}$ and $M_{\rm d}$ are
treated as variable parameters with time.  The above-mentioned set of
values of the parameters which are kept constant throughout the
numerical calculations secures positive mass density everywhere and free
of singularities.

For simplicity, we assume that mass is transported from the disk to the
nucleus, in such as way, that we have a linear increase in the mass of
the nuclear core, while a simultaneous linear decrease in the mass of
the disk takes place.  The linear mass transportation follows the
equations
\begin{eqnarray}
M_{\rm n} = M_{\rm n0} + k \ t, \nonumber\\
M_{\rm d} = M_{\rm d0} - k \ t,
\label{trans}
\end{eqnarray}
where $M_{\rm n0}$ and $M_{\rm d0}$ are the initial values of the mass
of the nucleus and the disk, respectively, while $k > 0$ is the
proportionality constant which determines the timescale of the galactic
evolution.  We choose $M_{\rm n0} = 0$, $M_{\rm d0} = 7500$ and $k =
0.05$, so at the beginning of the galactic evolution $(t = 0)$ there is
no nucleus formed and all the mass is concentrated in the disk.  We also
assume that the linear rate described by Eqs.  (\ref{trans}) is slow
compared to the orbital period of the stars and is therefore adiabatic.
This is true because the mass transportation lasts for $10^{4}$ time
units, while the orbital period is about two orders of magnitude
smaller.  Furthermore, the mass transportation and the galactic
evolution stops after $10^{4}$ time units when the final value of the
mass of the nucleus and that of the disk is $M_{\rm n} = 500$ and
$M_{\rm d} = 7000$, respectively.  The particular final values of the
parameters were chosen with a Milky Way-type galaxy in mind (e.g.,
\citealt{AS91}).  It is well known, that such mass transportation
mechanisms are usually met in the central regions of active galaxies and
they are responsible for the enormous luminosity of the quasars which
are hosted in the active galactic nuclei of the galaxies (see e.g.,
\citealt{CZ99}).  This is the main reason why we refer to the central mass
concentration as a ``nucleus" rather than as bulge (see also
\citealt{Z12}).

\begin{figure}[!tH]
\begin{center}
\includegraphics[width=0.7\hsize]{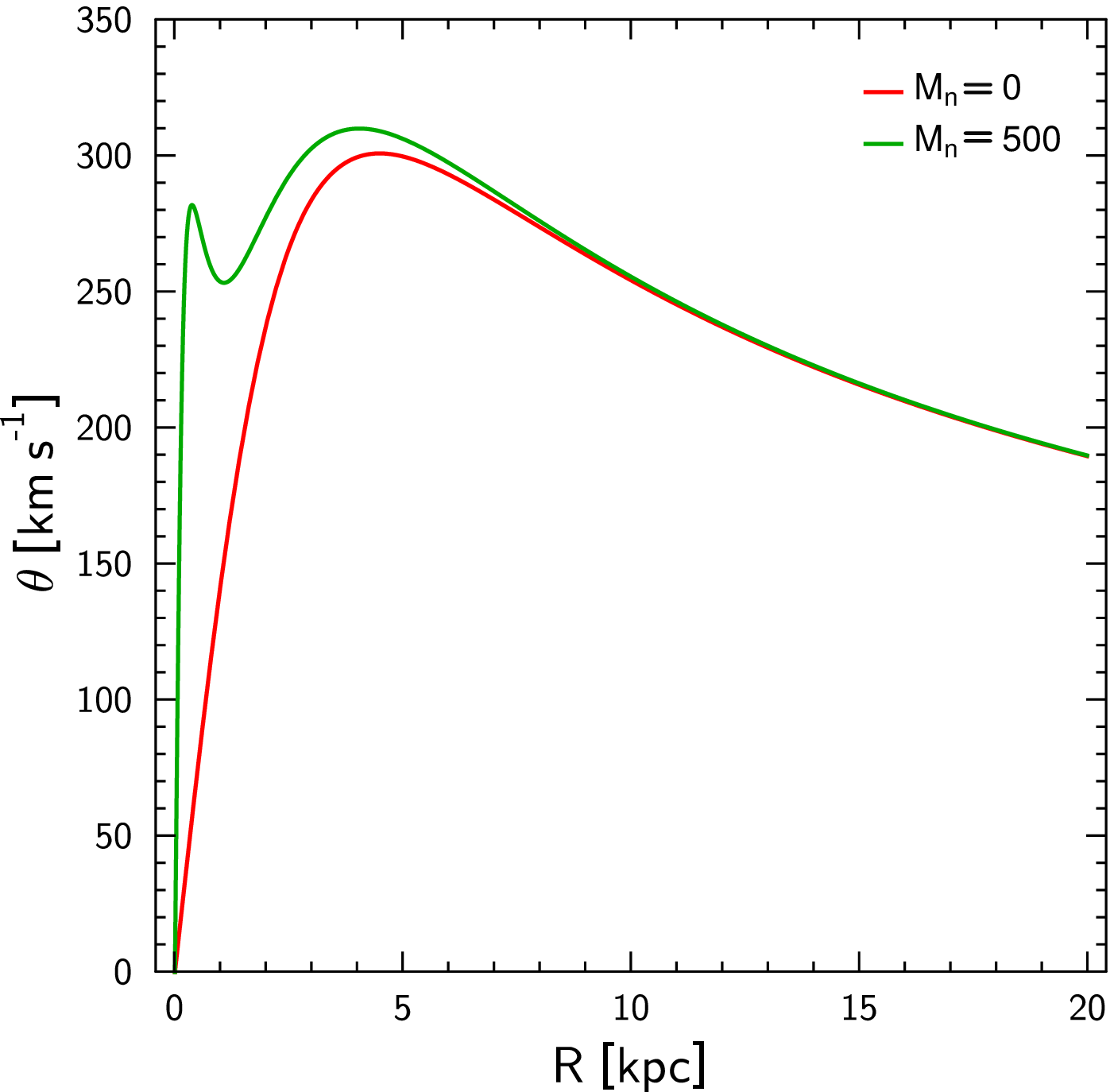}
\end{center}
\captionb{1}{A plot of the total  circular velocity of our galactic
model when $M_{\rm n} = 0$ and  $M_{\rm d} = 7500$ (red) and when
$M_{\rm n} = 500$ and $M_{\rm d} = 7000$ (green).}
\label{rotvel}
\end{figure}

The circular velocity in the galactic plane where $z = 0$ (see e.g.,
\citealt{Z11}) is undoubtedly one of the most important physical
quantities in disk galaxies and can be calculated as
\begin{equation}
\theta(R) = \sqrt{R\left|\frac{\partial \Phi(R,z)}{\partial R}\right|_{z = 0}}.
\label{rcur}
\end{equation}
In Fig.  \ref{rotvel} we present a plot of $\theta(R)$ at the beginning
$(t = 0)$ and at the end $(t = 10000)$ of the galactic evolution.  The
red line corresponds to the case where $M_{\rm n} = 0$ and $M_{\rm d} =
7500$, that is when the central nucleus is absent, while the green line
corresponds to the case where $M_{\rm n} = 500$ and $M_{\rm d} = 7000$
and a fully developed dense, massive nucleus lies in the galactic core.
It is seen that at relatively low galactocentric distances ($R \leq 5$
kpc) the pattern of the two curves is completely different, while at
large distances from the galactic center ($R > 5$ kpc), on the other
hand, both curves fully coincide.  We also observe the characteristic
local minimum (at $R \simeq 1$ kpc) of the rotation curve when the
massive nucleus is present, which appears when fitting observed data to
a galactic model (e.g., \citealt{GHBL10,IWTS13}).

Exploiting the fact that the $L_z$-component  of the total angular
momentum is conserved because the gravitational  potential $\Phi(R,z)$
is axially symmetric, orbits can be described by   means of the
effective potential
\begin{equation}
\Phi_{\rm eff}(R,z) = \Phi(R,z) + \frac{L_z^2}{2R^2}.
\label{veff}
\end{equation}

In this case, the basic equations of motion on the meridional plane take
the form
\begin{equation}
\ddot{R} = - \frac{\partial \Phi_{\rm eff}}{\partial R}, \ \ \ \ddot{z} = - \frac{\partial \Phi_{\rm eff}}{\partial z},
\label{eqmot}
\end{equation}
while the equations governing the evolution of a deviation vector
${\bf{w}} = (\delta R, \delta z, \delta \dot{R}, \delta \dot{z})$, which
joins the corresponding phase space points of two initially nearby
orbits, needed for the calculation of the standard indicators of chaos
(the SALI in our case), are given by the following variational equations
\begin{eqnarray}
\dot{(\delta R)} &=& \delta \dot{R}, \ \ \ \dot{(\delta z)} = \delta \dot{z}, \nonumber \\
(\dot{\delta \dot{R}}) &=&
- \frac{\partial^2 \Phi_{\rm eff}}{\partial R^2} \delta R
- \frac{\partial^2 \Phi_{\rm eff}}{\partial R \partial z}\delta z,
\nonumber \\
(\dot{\delta \dot{z}}) &=&
- \frac{\partial^2 \Phi_{\rm eff}}{\partial z \partial R} \delta R
- \frac{\partial^2 \Phi_{\rm eff}}{\partial z^2}\delta z.
\label{vareq}
\end{eqnarray}

Consequently, the  corresponding Hamiltonian to the effective potential
given in Eq. (\ref{veff}) can be written as
\begin{equation}
H = \frac{1}{2} \left(\dot{R}^2 + \dot{z}^2 \right) + \Phi_{\rm eff}(R,z) = E,
\label{ham}
\end{equation}
where $\dot{R}$ and $\dot{z}$ are momenta per unit mass, conjugate to
$R$ and $z$, respectively, while $E$ is the numerical value of the
Hamiltonian, which is conserved only in the time-independent model.
Therefore, an orbit is restricted to the area in the meridional plane
satisfying $E \geq \Phi_{\rm eff}$, while all other regions are
forbidden to the star.

\sectionb{3}{COMPUTATIONAL METHODS}
\label{cometh}

\begin{figure}[!tH]
\begin{center}
\includegraphics[width=0.7\hsize]{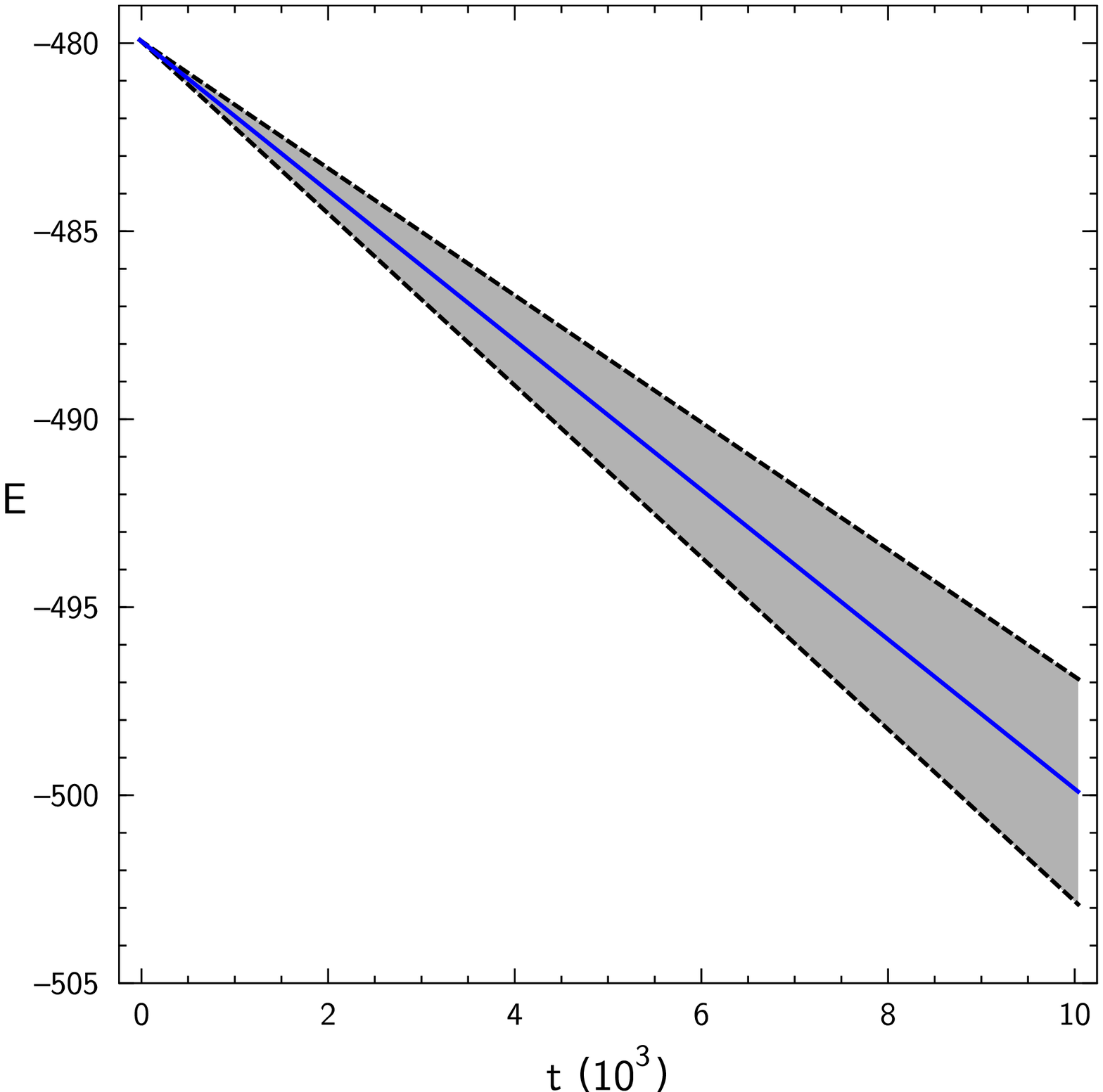}
\end{center}
\captionb{2}{Evolution of the value of the time-dependent Hamiltonian as
a function of time for 10000 orbits initiated at the phase $(R,\dot{R})$
space with $E_0 = -480$.  The blue line indicates the average
Hamiltonian change, while the black, dashed lines delimit the
boundaries of the energy variation $dE$.}
\label{ener}
\end{figure}

In order to examine the orbital dynamics (regularity or chaoticity) of
the galaxy model, we need to establish some samples of initial
conditions of orbits.  The best approach, undoubtedly, would be
to extract these samples of orbits from the distribution function of
the model potential.  Unfortunately, this is not available, so we
followed another course of action.  To determine the character of the
orbits in our model, we chose dense grids of initial conditions in the
phase $(R,\dot{R})$ space regularly distributed in the area allowed by
the value of the value of the Hamiltonian $E$.  At this point we have to
point out that the value of the energy controls the grid size
and
particularly the $R_{\rm max}$ which is the maximum possible value of
the $R$ coordinate.  On this basis, we chose as initial energy level the
value $E_0 = -480$ which yields $R_{\rm max} \simeq 15$ kpc.  Moreover,
the angular momentum of all orbits is $L_{\rm z} = 10$ and remains
constant throughout.  Since the Hamiltonian of our model is
time-dependent, its value is {\it not} conserved during the galactic
evolution.  To determine the change in the value of the Hamiltonian,
i.e. $dE = (E - E_0)/E_0$, we chose 10000 random initial conditions
in the phase space with $E_0 = -480$ and we recorded the value of the
Hamiltonian of every orbit during the galactic evolution until $t =
10^4$ time units when the mass transportation stops.  Our results are
presented in Fig.  \ref{ener} where we observe the evolution of the
value of Hamiltonian as a function of time.  It was found that for all
integrated orbits their energy is linearly reduced with time within the
gray-shaded area of the plot.  Furthermore, the change of the value of
the Hamiltonian ranges from about 3.5\% to 4.8\% and the black, dashed
lines in Fig.  \ref{ener} indicate the corresponding minimum and maximum
linear trends.

Taking into account that the range of values regarding the change of the
Hamiltonian is relatively small, we decided to adopt an average linear
change for all orbits in our model.  Our calculations show that the
average Hamiltonian change, shown in blue in Fig.  \ref{ener}, follows
the linear rule
\begin{equation}
E = E_0 - q \ t,
\label{dEt}
\end{equation}
where $E_0 = -480$, while $q = 0.002$ so that $E = -500$ at $t = 10^4$
time units.  Relation (\ref{dEt}) is a powerful tool because it allows
us to know the exact value of the Hamiltonian at every time step of the
galactic evolution.  Now we can use this valuable information in order
to explore the orbital structure of the galaxy at various time points
during the mass transportation.  In particular, since we have equations
describing the time-evolution of $M_{\rm n}$, $M_{\rm d}$ and $E$, we
can pick some time points $t_i$, freeze the mass transportation and use
the time-independent model in order to integrate and classify the
orbits, following the numerical approach used in \citet{MBS13}.
Therefore, for each set of values of $M_{\rm n}(t_i)$  and $M_{\rm
d}(t_i)$, we define a dense grid of initial conditions in the phase space
regularly distributed in the area allowed by the corresponding value of
the Hamiltonian $E(t_i)$.  The step separation of the initial conditions
along the $R$ and $\dot{R}$ axes (or, in other words, the density of the
grid) was controlled in such a way that always there are at least 50000
orbits to be integrated and classified.  The grids of initial conditions
of orbits whose properties will be examined are defined as follows:  we
consider orbits with initial conditions $(R_0, \dot{R_0})$ with $z_0 =
0$, while the initial value of $\dot{z_0}$ is obtained from the
Hamiltonian (\ref{ham}).  For each initial condition, we numerically
integrated the equations of motion (\ref{eqmot}) as well as the
variational equations (\ref{vareq}) with a double precision
Bulirsch-Stoer \verb!FORTRAN 77! algorithm (e.g., \citealt{PTVF92}) with
a small time step of the order of $10^{-2}$, which is sufficient enough
for the desired accuracy of our computations (i.e., our results
practically do not change by halving the time step).  In all cases, the
energy integral\footnote{Remember that these orbits are integrated in
the frozen time-independent model, where the energy integral is valid.}
(Eq.  \ref{ham}) was conserved better than one part in $10^{-11}$,
although for most orbits it was better than one part in $10^{-12}$.

For determining the regular or chaotic nature of orbits, we chose the
SALI indicator \citet{S01}.  The time-evolution of SALI strongly depends
on the nature of the computed orbit since when an orbit is regular the
SALI exhibits small fluctuations around non-zero values, while on the
other hand, in the case of a chaotic orbit, the SALI after a small
transient period tends exponentially to zero approaching the limit of
the accuracy of the computer $(10^{-16})$.  Therefore, the particular
time-evolution of the SALI allow us to distinguish fast and safely
between regular and chaotic motion.  Nevertheless, we have to define a
specific numerical threshold value for determining the transition from
regularity to chaos.  After conducting extensive numerical experiments,
integrating many sets of orbits, we conclude that a safe threshold value
for the SALI, taking into account the total integration time of $10^4$
time units, is the value $10^{-7}$.  In order to decide whether an orbit
is regular or chaotic, one may follow the usual method according to
which we check, after a certain and predefined time interval of
numerical integration, if the value of SALI has become less than the
established threshold value.  Therefore, if SALI $\leq 10^{-7}$ then the
orbit is chaotic, while if SALI $ > 10^{-7}$ then the orbit is regular.
However, depending on the particular location of each orbit, this
threshold value can be reached more or less quickly, as there are
phenomena that can hold off the final classification of an orbit (i.e.,
there are special orbits called ``sticky" orbits, which behave regularly
for long time periods before they finally drift away from the boundaries
of regular regions and start to wander in the chaotic domain, thus
revealing their true chaotic nature fully.  For the computation of SALI
we used the \verb!LP-VI! code \citet{CMD14}, a fully operational routine
which efficiently computes a suite of many chaos indicators for
dynamical systems in any number of dimensions.

Each orbit in the frozen time-independent Hamiltonian was numerically
integrated for a time interval of $10^4$ time units ($10^{12}$ yr),
which corresponds to a time span of the order of hundreds of orbital
periods.  The particular choice of the total integration time is an
element of great importance, especially in the case of the sticky
orbits.  A sticky orbit could be easily misclassified as regular by any
chaos indicator\footnote{~Generally, dynamical methods are broadly split
into two types:  (i) those based on the evolution of sets of deviation
vectors to characterize an orbit and (ii) those based on the frequencies
of the orbits that extract information about the nature of motion only
through the basic orbital elements without the use of deviation
vectors.}, if the total integration interval is too small, so that the
orbit does not have enough time to reveal its true chaotic character.
Thus, all the initial conditions of the orbits of a given grid were
integrated, as we already said, for $10^4$ time units, thus avoiding
sticky orbits with a stickiness at least of the order of $10^2$ Hubble
times.  All the sticky orbits that do not show any signs of chaoticity
for $10^4$ time units are counted as regular orbits since such vast
sticky periods are completely out of the scope of our research.

A vital clarification regarding the nomenclature of the orbits should be
made before closing this Section.  All orbits of an axially symmetric
potential are in fact three-dimensional (3D) loop orbits, i.e., orbits
that always rotate around the axis of symmetry in the same direction.
However, in dealing with the meridional plane, the rotational motion is
lost, so the path that the orbit follows onto this plane can take any
shape, depending on the nature of the orbit.  Following the same
approach of the previous papers of this series, we characterize an
orbit according to its behavior in the meridional plane.  If, for
example, an orbit is a rosette lying in the equatorial plane of the
axisymmetric potential, it will be a linear orbit in the meridional
plane, a tube orbit it will be a 2:1 orbit, etc.  We should emphasize
that we use the term ``box orbit" for an orbit that conserves
circulation, but this refers {\it only} to the circulation provided
by the meridional plane itself.  Because of the their boxlike shape in
the meridional plane, such orbits were originally called ``boxes" (e.g.,
\citealt{O62}), even though their three-dimensional shapes are more
similar to doughnuts (more details can be found in the review of
\citealt{M99}).  Nevertheless, we kept this formalism to maintain
continuity with all the previous papers of this series.

\sectionb{4}{NUMERICAL RESULTS}
\label{numres}

\begin{figure*}[!tH]
\begin{center}
\resizebox{0.7\hsize}{!}{\includegraphics{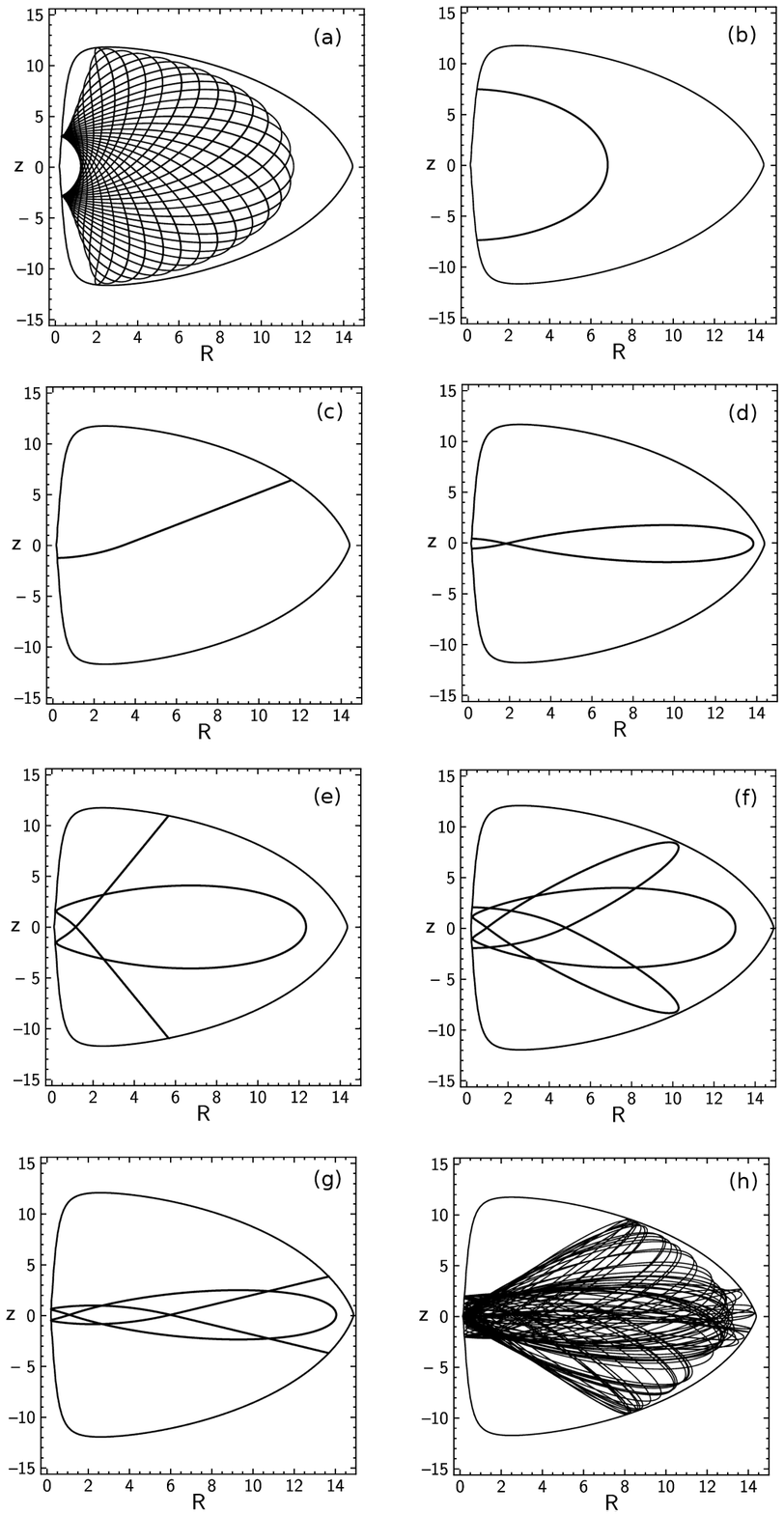}}
\end{center}
\captionb{3}{Orbit collection of the basic  types of orbits in our
galaxy model: (a) box orbit; (b) 2:1  banana-type orbit; (c) 1:1 linear
orbit; (d) 2:3 fish-type orbit; (e)  4:3 boxlet orbit; (f) 6:5 boxlet
orbit; (g) 4:5 boxlet orbit, one of our ``orbits
with other resonance"; (h) chaotic orbit.}
\label{orbs}
\end{figure*}

\begin{table}
   \captiontb{1}{Types and initial  conditions of the orbits shown in
Fig. \ref{orbs}(a-h). In all cases,  $z_0 = 0$, $\dot{z_0}$ is found
from the Hamiltonian, Eq. (\ref{ham}),  while $T_{\rm per}$ is the
period of the resonant parent periodic orbits.}
   \label{table}
   \begin{center}
   \begin{tabular}{@{}|c|c|c|c|c|}
      \hline
      Figure & Type & $R_0$ & $\dot{R_0}$ & $T_{\rm per}$  \\
      \hline
      \ref{orbs}a &  box        &  1.20000000 &  0.00000000 &          - \\
      \ref{orbs}b & 2:1 banana  &  6.94608753 &  0.00000000 & 1.82936813 \\
      \ref{orbs}c & 1:1 linear  &  3.48802471 & 38.08701776 & 1.52281338 \\
      \ref{orbs}d & 2:3 boxlet  & 14.12356085 &  0.00000000 & 3.07990781 \\
      \ref{orbs}e & 4:3 boxlet  & 12.62569659 &  0.00000000 & 5.95238344 \\
      \ref{orbs}f & 6:5 boxlet  & 13.27429077 &  0.00000000 & 9.51868683 \\
      \ref{orbs}g & 4:5 boxlet  & 14.30689514 &  0.00000000 & 6.47237905 \\
      \ref{orbs}h & chaotic     &  0.12000000 &  0.00000000 &          - \\
      \hline
   \end{tabular}
\end{center}
\end{table}

This Section contains the main numerical results of both the
time-independent and the time-evolving model.  In particular, we
numerically integrate several sets of orbits, in an attempt to determine
the regular or chaotic nature of motion of stars.  First we begin with
the time-independent model and use the sets of initial conditions
described in the previous section to construct the respective grids,
always adopting values inside the limiting curve.  In all cases, the
initial value of the energy was set equal to $-480$, while the angular
momentum of the orbits is $L_{\rm z} = 10$.  To study how the mass
transportation from the disk to the central nucleus influences the level
of chaos, we choose representative time points of the galactic evolution
such as $t_i = \{0,2000,4000, ..., 10000\}$.  Once the exact values of
the parameters $M_{\rm n}$, $M_{\rm d}$ and $E$ are known through Eqs.
(\ref{trans}) and (\ref{dEt}), we compute a set of initial conditions as
described in Section \ref{cometh} and we integrate the corresponding
orbits computing the SALI of the orbits and then classifying regular
orbits into different families.

Our numerical investigation suggests that in our galaxy model there are
eight basic types of orbits:  (i) chaotic orbits; (ii) box orbits; (iii)
1:1 linear orbits; (iv) 2:1 banana-type orbits; (v) 2:3 fish-type
orbits; (vi) 4:3 resonant orbits; (vii) 6:5 resonant orbit and (viii)
orbits with other secondary resonances (i.e., all resonant orbits not
included in the former categories).  It turns out that for each resonant
family included in the ``other category" the corresponding percentage is
less than 1\% in all cases, and therefore their contribution to the
overall orbital structure of the galaxy is practically insignificant.
The $n:m$ notation\footnote{~A $n:m$ resonant orbit would be represented
by $m$ distinct islands of invariant curves in the $(R,\dot{ R})$ phase
plane and $n$ distinct islands of invariant curves in the $(z,\dot{z})$
surface of section.} we use for the regular orbits is according to
\citet{CA98} and \citet{ZC13}, where the ratio of those integers
corresponds to the ratio of the main frequencies of the orbit, where
main frequency is the frequency of greatest amplitude in each
coordinate.  Main amplitudes, when having a rational ratio, define the
resonances of an orbit.  In Fig.  \ref{orbs}(a-h) we present examples of
each of the basic types of regular orbits, plus an example of a chaotic
one.  In all cases, we set $M_{\rm n} = 500$ and $E = -500$ (except for
the 4:5 and 6:5 resonant orbits, where $M_{\rm n} = 100$ and $E =
-484$).  The orbits shown in Figs.  \ref{orbs}a and \ref{orbs}h were
computed until $t = 100$ time units, while all the parent periodic
orbits \citet{Z13} were computed until one period has completed.  The
outermost black curve circumscribing each orbit is the limiting Zero
Velocity Curve (ZVC) in the meridional plane which is defined as
$\Phi_{\rm eff}(R,z) = E$.  Table \ref{table} shows the types and the
initial conditions for each of the depicted orbits; for the resonant
cases, the initial conditions and the period $T_{\rm per}$ correspond to
the parent\footnote{~For every orbital family there is a parent (or
mother) periodic orbit, that is, an orbit that describes a closed
figure.  Perturbing the initial conditions which define the exact
position of a periodic orbit we generate quasi-periodic orbits that
belong to the same orbital family and librate around their closed parent
periodic orbit.} periodic orbits.

We would like to point out at this point, that the 1:1 resonance is
usually the hallmark of the loop orbits and both coordinates oscillate
with the same frequency in their main motion.  Their mother orbit is a
closed loop orbit.  Moreover, when the oscillations are in phase, the
1:1 orbit degenerates into a linear orbit (the same as in Lissajous
figures made with two oscillators).  In our meridional plane, however,
1:1 orbits do not have the shape of a loop.  In fact, their mother orbit
is linear (e.g., Fig.  \ref{orbs}c), and thus they do not have a hollow
(in the meridional plane), but fill a region around the linear mother,
always oscillating along the $R$ and $z$ directions with the same
frequency.  We designate these orbits ``1:1 linear open orbits" to
differentiate them from true meridional plane loop orbits, which have a
hollow and also always rotate in the same direction.

\begin{figure*}[!tH]
\begin{center}
\resizebox{0.8\hsize}{!}{\includegraphics{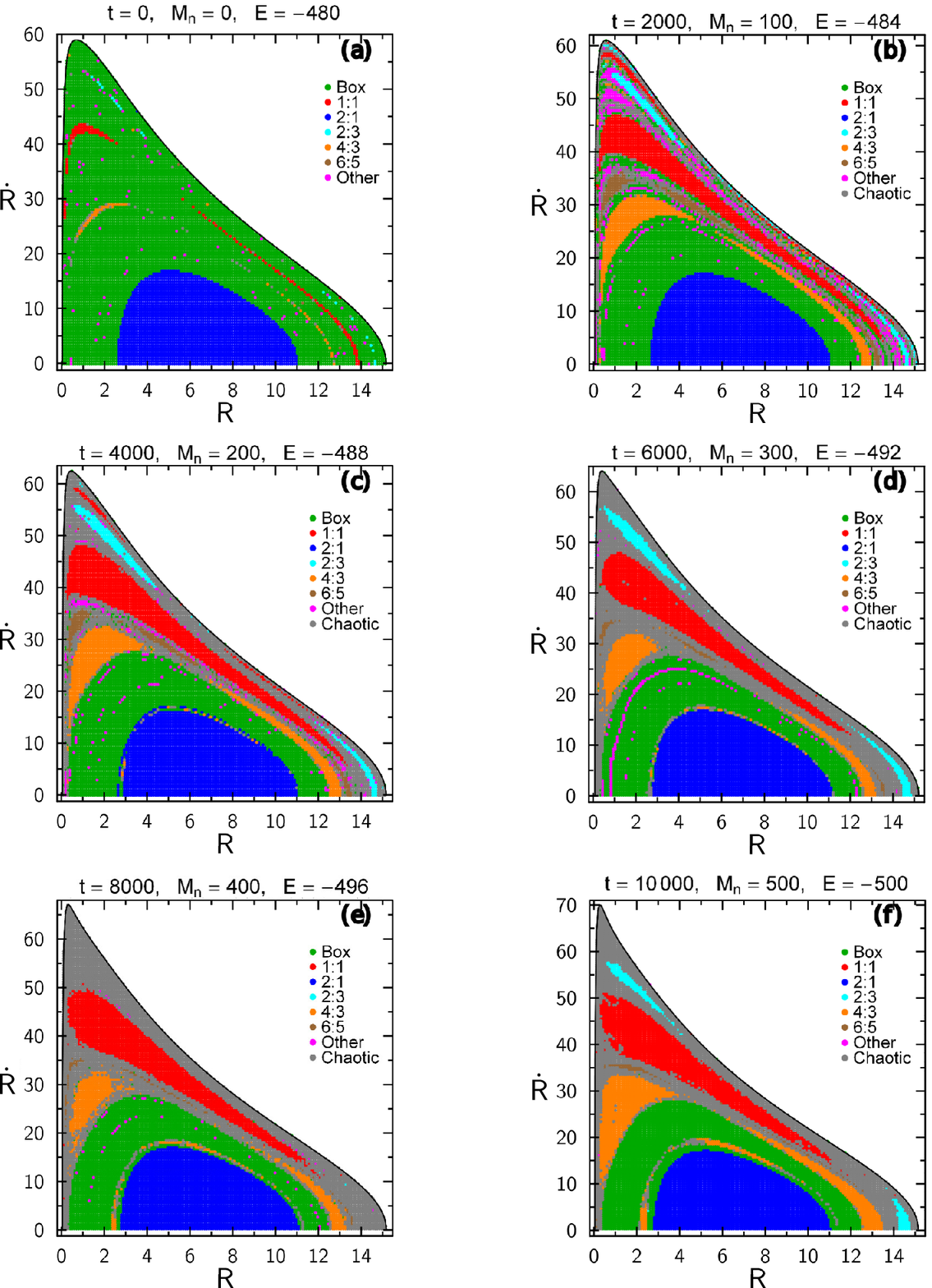}}
\end{center}
 \captionb{4}{Orbital structure of the phase $(R,\dot{R})$ plane of our
galaxy model at several time points of the galactic evolution.}
\label{grd}
\end{figure*}

In the following Figs.  \ref{grd}(a-f) we present six grids of initial
conditions $(R_0,\dot{R_0})$ of orbits that we have classified for
different values of the final mass of the nucleus $M_{\rm n}$ in the
frozen time-independent model.  These color-coded grids of initial
conditions are equivalent to the classical Poincar\'{e} Surfaces of
Section (PSS) and allow us to determine what types of orbits occupy
specific areas in the phase $(R,\dot{R})$ plane and also to follow the
changes that orbits undergo.  The outermost black thick curve is the
limiting curve which is defined as \begin{equation} \frac{1}{2}
\dot{R}^2 + \Phi_{\rm eff}(R,z = 0) = E. \label{zvc} \end{equation} In
Fig.  \ref{grd}a which corresponds to time point $t = 0$ where the
central nucleus is absent $(M_{\rm n} = 0)$ we observe that the entire
phase plane is covered by initial conditions of regular orbits, while
chaotic motion, if any, is negligible.  In particular, initial
conditions of box orbits occupy the vast majority of the phase space, a
well-formed island of 2:1 resonant orbits is present at the center of
the grid, while there are also several smaller stability islands of
other resonant orbits embedded in the extended box area.  The structure
of the phase plane however, changes drastically in Fig.  \ref{grd}b at
$t = 2000$ time units when $M_{\rm n} = 100$.  It is seen that the box
area is significantly reduced and especially at the outer parts of the
phase plane there is a strong presence of stability islands of resonant
orbits surrounded by a thin chaotic layer.  It is interesting to note
that inside this chaotic layer one may identify several sets of tiny
stability islands corresponding to higher resonant orbits which in our
classification belong to the ``other" category.  The area on the phase
plane occupied by these secondary higher resonant orbits is considerably
confined when $t = 4000$ time units and $M_{\rm n} = 200$.  Indeed we
see in Fig.  \ref {grd}c that the corresponding stability islands are
hardly visible, while at the same time the chaotic zone is amplified.
The same pattern continues in Fig.  \ref{grd}d for $t = 6000$ time units
and $M_{\rm n} = 300$ where the chaotic area is more prominent, while
some thin filaments of initial conditions of secondary resonant inside
the box region are observed.  The mass transport is still in progress
and for $t = 8000$ time units which corresponds to $M_{\rm n} = 400$
there is a major difference in the phase space, that is the absence of
the 2:3 stability islands.  Additional numerical calculations reveal
that for $M_{\rm n} = 400$ the 2:3 resonance is unstable.  This means
that the periodic point of the 2:3 resonance is indeed present in Fig.
\ref{grd}e, although evidently deeply buried in the chaotic domain.
Finally, in Fig.  \ref{grd}f, where $t = 10000$ time units and $M_{\rm
n} = 500$, we observe that the 2:3 stability islands reappear in the
phase space.

Looking carefully the color-coded grids presented in Figs.\ref{grd}(a-f)
we can distinguish the location of the seven main types of regular
orbits discussed earlier:  (i) 2:1 banana-type orbits located to the
central region of the grid; (ii) box orbits situated mainly outside of
the 2:1 resonant orbits; (iii) 1:1 open linear orbits form the elongated
island of initial conditions; (iv) 2:3 fish-type resonant orbits form
the set of three small islands\footnote{~It should be pointed out that
the color-coded grids of Fig.  \ref{grd}(a-f) show only the $\dot{R} >
0$ part of the phase plane; the $\dot{R} < 0$ is symmetrical.
Therefore, in many resonances not all the corresponding stability
islands are present (e.g., for the 2:3 and 4:3 resonances only two
(actually one and a half) of the three islands are shown).} at the
outer parts of the grid; (v) 4:3 resonant orbits form the chain of three
islands; (vi) 6:5 resonant orbits producing five small stability islands
inside the chaotic region and (vii) other types of resonances producing
extremely small islands embedded both in the chaotic and box areas.
Chaos on the other hand, is mainly confined only at the outer parts of
the phase plane.  It is evident, that as the central nucleus becomes
more and more massive gaining mass from the galactic disk, the amount of
chaos in the phase space increases.  It should also be mentioned that
the allowed radial velocity $\dot{R}$ of the stars near the center of
the galaxy is increasing during the galactic evolution, where the mass
of the nucleus also increases.

\begin{figure}[!tH]
\begin{center}
\includegraphics[width=0.7\hsize]{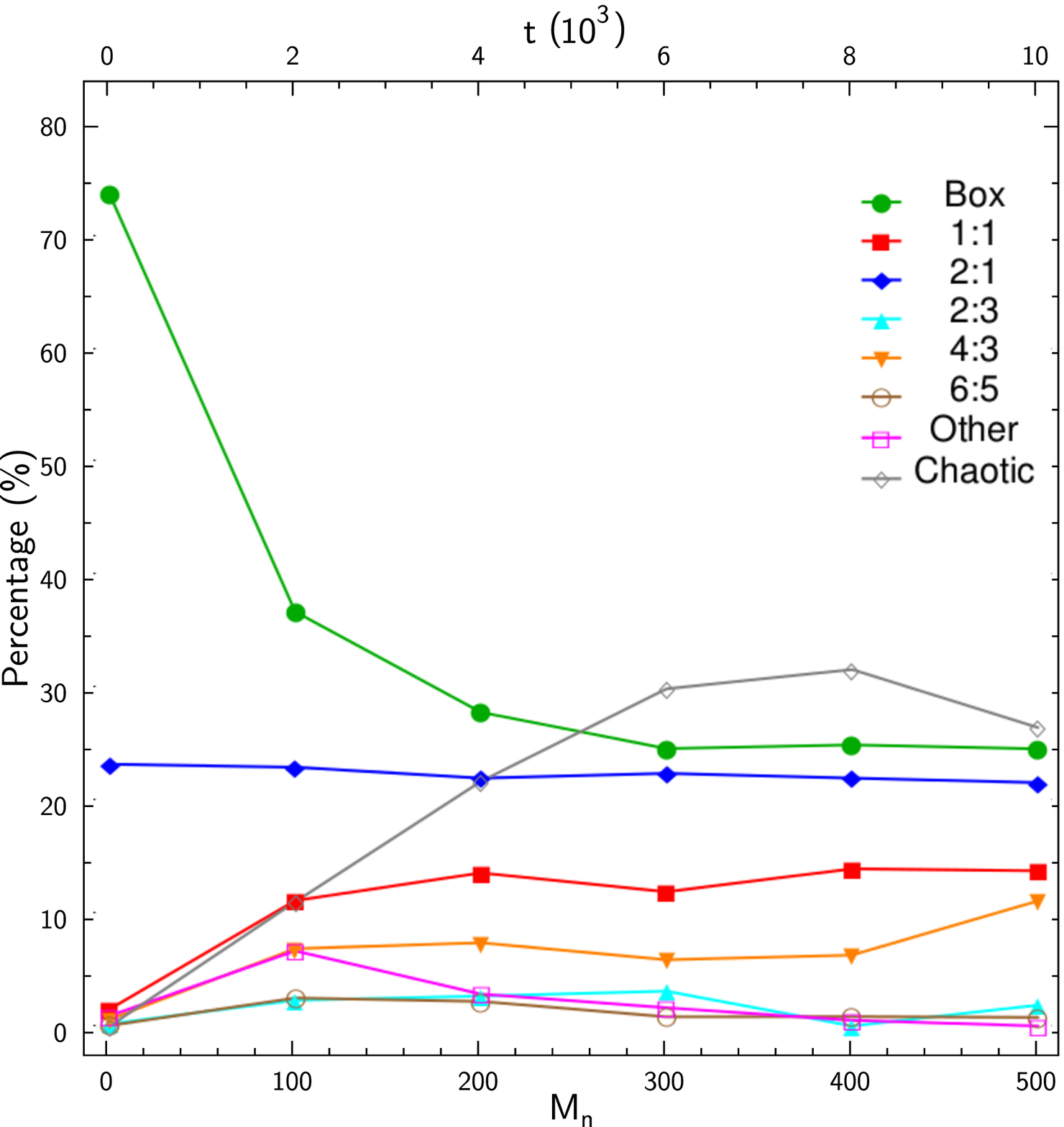}
\end{center}
\captionb{5}{Time-evolution of the percentages of the  different types
of orbits in the phase $(R,\dot{R})$ space of our  galaxy model, as a
function of the final mass of the nucleus $M_{\rm n}$.}
\label{percs}
\end{figure}

The time-evolution of the percentages of the chaotic and all types of
regular orbits as a function of the final mass of the central nucleus
$M_{\rm n}$ is presented in Fig.  \ref{percs}.  One may observe that at
the beginning of the galactic evolution $(t = 0)$ when the central
nucleus is absent, there is no chaos whatsoever, and more than 70\% of
the phase space is dominated by initial conditions of box orbits.
However, as mass begins to be transferred from the disk to the core,
box
orbits are depleted and this transportation starts to trigger chaotic
motion.  In particular, we see that as the nucleus becomes more and more
massive with time the percentage of box orbits is significantly reduced
and for $M_{\rm n} > 300$ it saturates around 25\%, while at the same
time the rate of chaotic orbits exhibits a considerable growth and for
about $t > 6000$ time units chaotic orbits is the most populated family,
although at high enough values of $M_{\rm n}$ their rate displays a
minor decrease.  At the end of the galactic evolution at $t = 10000$
time units, or in other words when $M_{\rm n} = 500$, the rates of box
and chaotic orbits tend to a common value (around 25\%), thus sharing
half of the phase plane.  Our numerical analysis suggests that the
percentages of the rest of orbital families change very little during
the galactic evolution.  Indeed it is seen that the 2:1 resonant orbits
(meridional bananas) are almost unperturbed by the shifting of the mass
of the central nucleus occupying about 22\% of the phase space
throughout.  Furthermore, the rates of the 1:1 and 4:3 resonant families
display a monotone behavior roughly around 10\% during the mass
transportation.  In addition, we may say that, in general terms, all the
other resonant families (i.e., the 2:3, 6:5 and ``other") possess
throughout very low percentages (always less than 5\%) thus, the growth
of the mass of the nucleus only shuffles the orbital content among them.
Therefore, taking into account all the above-mentioned analysis, we may
conclude that in the phase $(R,\dot{R})$ space the types of orbits that
are mostly influenced by the mass transportation are the box and chaotic
orbits.  It is also interesting to note that the time-evolution of the
percentages of the orbits is very similar to that given in Fig. 6 in
\citet{ZC13}, where the influence of the mass of the nucleus in the
time-independent version of same model was investigated.

\begin{figure}[!tH]
\begin{center}
\includegraphics[width=0.9\hsize]{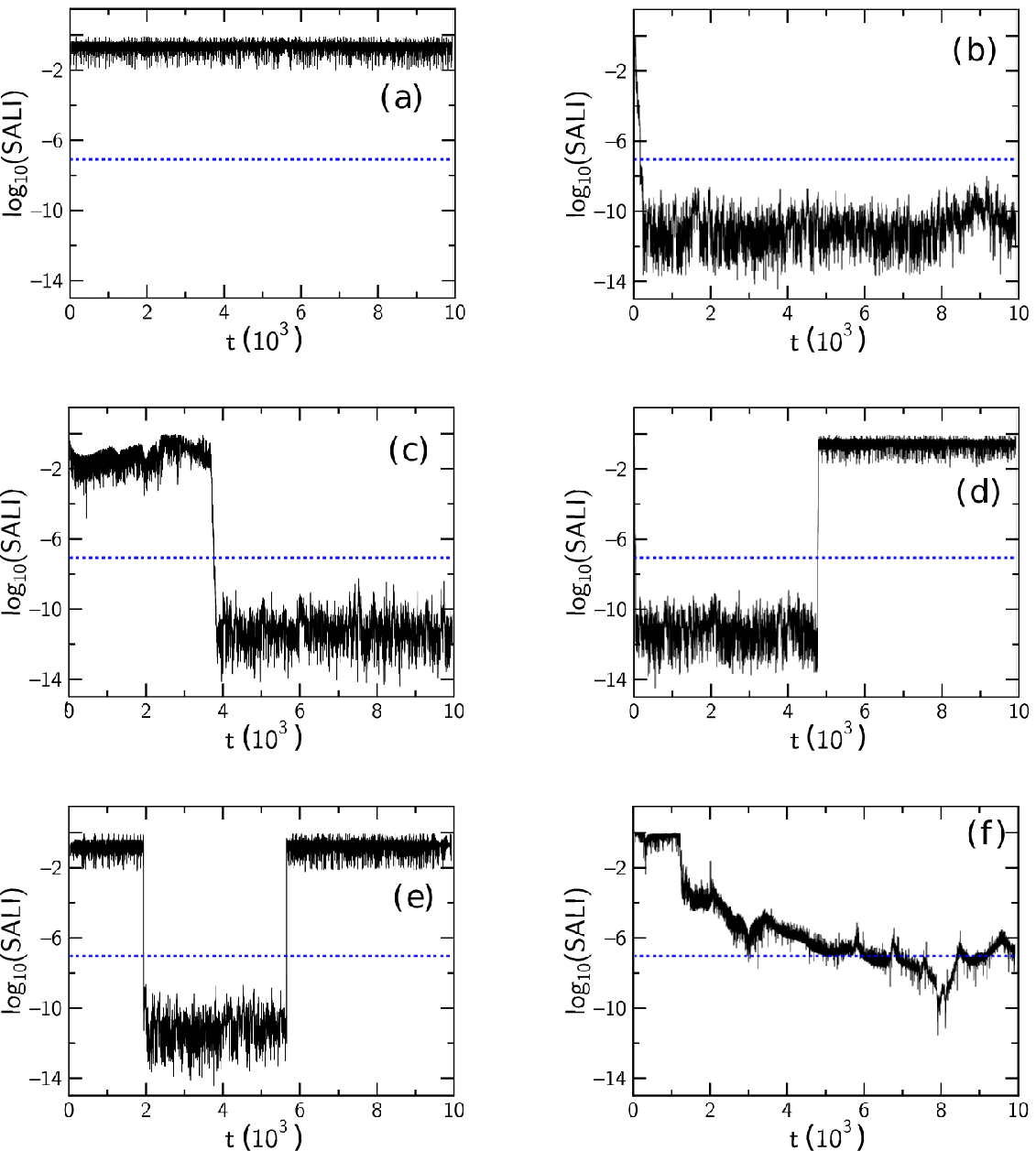}
\end{center}
\captionb{6}{Time-evolution of SALI for six different  types of orbits
in the time-dependent Hamiltonian. Details are given in the text.}
\label{salis}
\end{figure}

All the previous numerical results correspond to specific time points
off the galactic evolution.  As it was explained in the previous
Section, we can pick some characteristic time points, use the respective
values of the mass of the disk, the mass of the nucleus and the value of
the Hamiltonian and integrate orbits in the time-independent frozen
model.  Now we would like to investigate the character of orbits and the
transition from regularity to chaos and vice versa during the mass
transportation.  For this purpose, we use the full time-dependent
Hamiltonian as follows:  we define in the phase space a grid of 50000
initial conditions of orbits with $E_0 = -480$ and we vary the mass of
the nucleus form $M_{\rm n}(t_0 = 0) = 0$ to $M_{\rm n}(t_{\rm final} =
10000) = 500$ recording at each time step of the numerical integration
the value of SALI.  Our computations suggest that the nature of the
orbits can change either from ordered to chaotic and vice versa or not
change at all, as mass is transported from the disk and a massive
nucleus is developed in the central region of the galaxy.  In Fig.
\ref{salis}(a-f) we present the time-evolution of SALI of six different
orbits, as the total mass distribution of the galactic system changes
with time, following the set of equations (\ref{trans}).  For all six
orbits we set $z_0 = 0$, $\dot{R_0} = 0$, while $\dot{z_0}$ is obtained
from the initial value of the Hamiltonian which is $E_0 = -480$.  The
horizontal, blue, dashed line in Fig.  \ref{salis} corresponds to the
threshold value (SALI = $10^{-7}$) which separates regular from chaotic
motion.  Fig.  \ref{salis}a shows the time-evolution of SALI for an
orbit with $R_0 = 7.5$.  It is seen that the particular orbit starts as
regular and remains regular throughout the galactic evolution.  The
pattern of SALI shown in Fig.  \ref{salis}b on the other hand, suggests
that the orbit with $R_0 = 0.145$ maintains its chaotic character
regardless the mass transportation.  In Fig.  \ref{salis}c where $R_0 =
14.35$, we see that the orbit starts as regular but after about 3800
time units it becomes chaotic.  The exact opposite scenario is shown in
Fig.  \ref{salis}d where $R_0 = 0.23$.  This orbit exhibits chaotic
behavior for the first about 4800 time units of the galactic evolution
however, for $t > 4800$ it clearly becomes regular.  The transition from
regularity to chaos and vice versa may occur more than one time during
the mass transportation.  Fig.  \ref{salis}e shows such a characteristic
example of an orbit with $R_0 = 0.2217$ which display s chaotic nature
only in the interval $1950 < t < 5670$.  Nevertheless, the determination
of the character of an orbit is not always very easy.  This becomes
evident by inspecting Fig.  \ref{salis}f where one may observe that the
SALI of the orbit with $R_0 = 0.21$ oscillates around the threshold
value, thus preventing us from having a clear and definitive view
regarding the character of this orbit, which probably remains sticky
throughout the galactic evolution.  We must point out that these
dynamical transitions are not related by no means to stickiness or
ordinary diffusion phenomena that occur in time-independent systems.

\begin{figure}[!tH]
\begin{center}
\includegraphics[width=0.9\hsize]{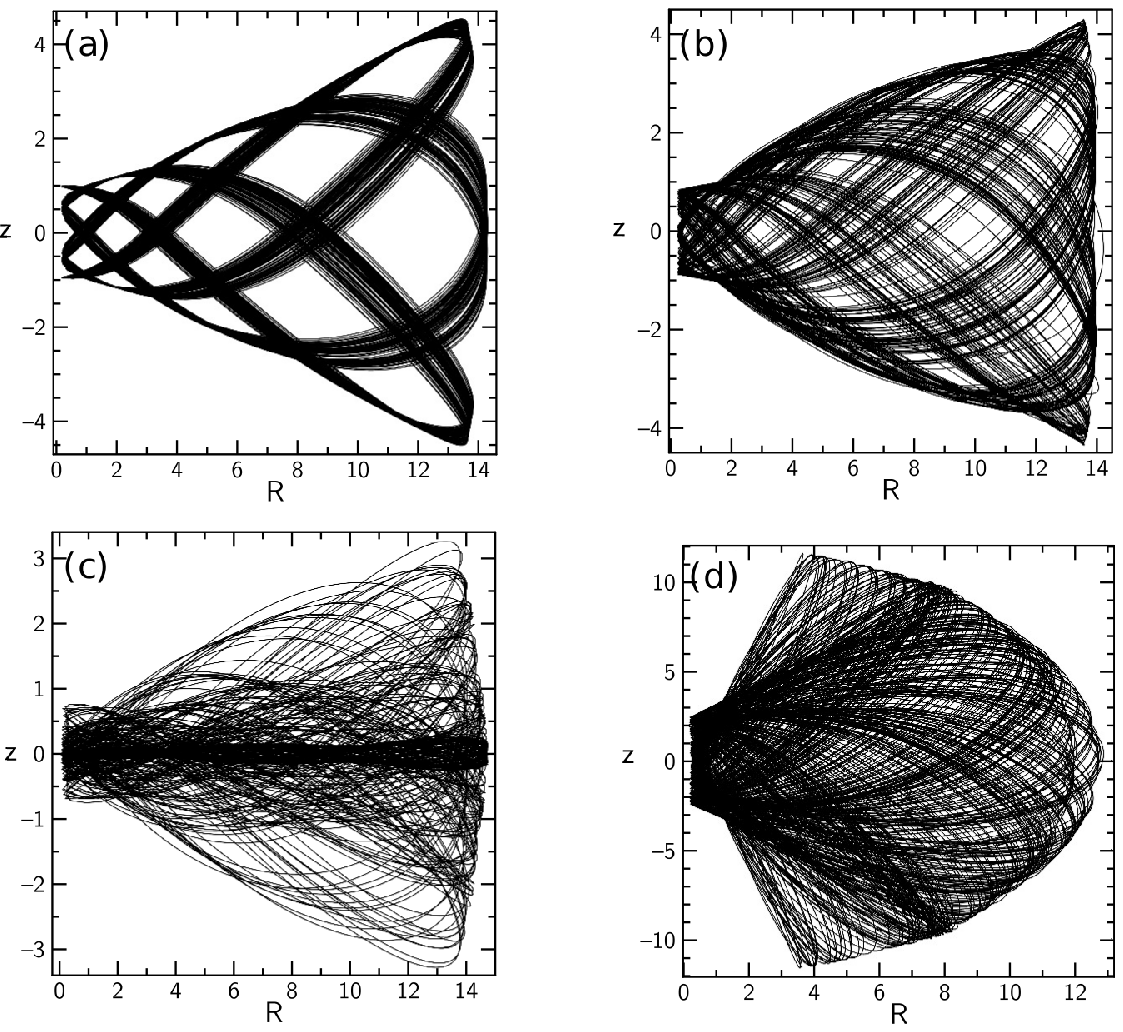}
\end{center}
\captionb{7}{The shape of the orbit discussed in Fig.  \ref{salis}c in
four time intervals of the galactic evolution.  (a-upper left):  $0 \leq
t \leq 500$; (b-upper right):  $3500 \leq t < 3700$; (c-lower left):
$3700 \leq t \leq 4100$; (d-lower right):  $8000 \leq t \leq 8500$.}
\label{orbtd}
\end{figure}

In order to have a more enlightening picture about the transition from
order to chaos during the mass transportation we provide in Fig.
\ref{orbtd}(a-d) the shape on the meridional $(R,z)$ plane of the orbit
explained in Fig.  \ref{salis}c for four time intervals of the galactic
evolution.  Fig.  \ref{orbtd}a shows the orbit for the first 500 time
units, where it is clearly seen that the orbit is beyond any doubt a
regular 6:7 resonant orbit.  In Fig.  \ref{orbtd}b where $3500 \leq t <
3700$ we observe that the path of the trajectory the star follows
becomes very unclear which is an indication of imminent chaotic motion.
Indeed in Fig.  \ref{orbtd}c where $3700 \leq t \leq 4100$ we see that
the transition from regularity to chaos has been completed, according to
the corresponding time-evolution of SALI which reported the transition
point around 3800 time units.  Moreover, it is observed in Fig.
\ref{orbtd}c that the test particle (star) spends a great deal of time
moving very close to the galactic plane.  Finally, in Fig.  \ref{orbtd}d
where $8000 \leq t \leq 8500$, that is an advanced stage of the galactic
evolution where the nucleus is massive enough, the complete chaotic
nature of the orbits is fully revealed and the star moves at relatively
high distances from the galactic plane up to about 10 kpc.  It should be
stressed out however, that the shape of an orbit gives only fast and
qualitative information which sometimes can be inconclusive or even
misleading regarding the nature of the orbit.  Therefore, only highly
accurate methods that use certain and objective numerical criteria, such
as the SALI, should be used for determining the characters of orbits.

\begin{figure}[!tH]
\begin{center}
\includegraphics[width=0.7\hsize]{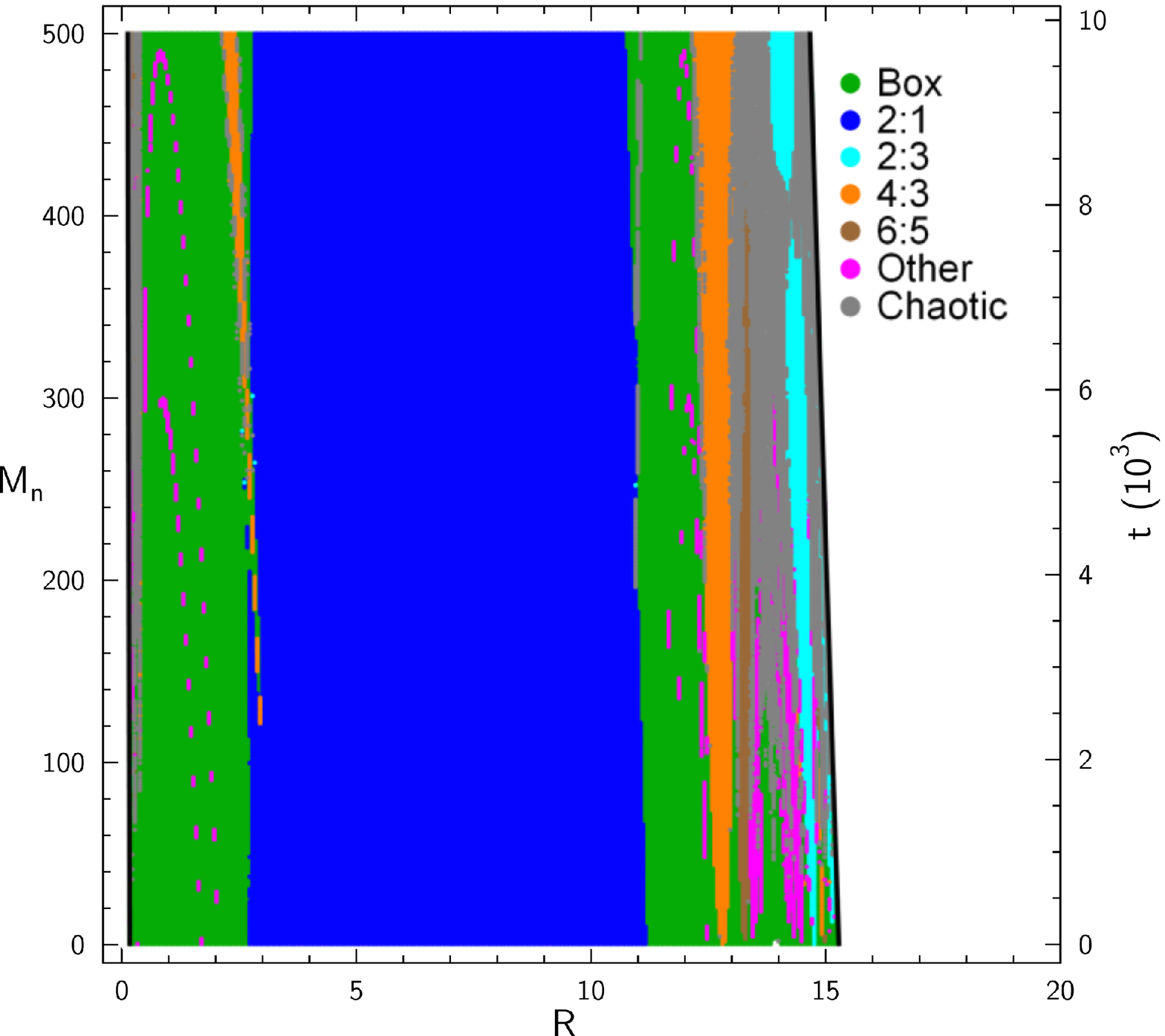}
\end{center}
 \captionb{8}{Orbital structure of the $(R,M_{\rm n})$-plane. This
 diagram gives us a detailed analysis of the evolution of orbits
 starting perpendicularly to the $R$-axis when the mass of the nucleus
varies in the interval $M_{\rm n} \in [0, 500]$ during
the galactic evolution for $0 \leq t \leq 10000$ time units.}
\label{Rt}
\end{figure}

The color-coded grids of initial conditions in the phase $(R,\dot{R})$
plane presented in Fig.  \ref{grd}(a-f) provide information on the phase
space mixing for only a specific time point of the galactic evolution
and for the corresponding value of the mass of the nucleus $M_{\rm n}$.
Following H\'{e}non's idea \citet{H69} however, we can consider a plane
which provides information about regions of regularity and regions of
chaos using the section $z = \dot{R} = 0$, $\dot{z} > 0$, i.e., the test
particles (stars) are launched on the $R$-axis, parallel to the $z$-axis
and in the positive $z$-direction.  Thus, in contrast to the previously
discussed grids (Fig.  \ref{grd}(a-f)), only orbits with pericenters on
the $R$-axis are included and, therefore, the value of $M_{\rm n}$ is
now used as an ordinate.  In this way, we can monitor how the mass of
the nucleus influences the overall orbital structure of our dynamical
system using a continuous spectrum of values of $M_{\rm n}$ rather than
a few discrete ones.  Fig.  \ ref{Rt} shows the orbital structure of the
$(R,M_{\rm n})$-plane, when $M_{\rm n} \in [0, 500]$ and $0 \leq t \leq
10000$ time units.  In order to be able to monitor with sufficient
accuracy and details the evolution of the families of orbits, we defined
a dense grid of $10^5$ initial conditions in the $(R,M_{\rm n})$-plane.
For creating this plane we use the time-independent model with the
values of $M_{\rm n}$, $M_{\rm d}$ and $E$ according to Eqs.
(\ref{trans}) and (\ref{dEt}).  It is evident, that the vast majority of
the grid is covered either by box or 2:1 resonant orbits, while initial
conditions of chaotic orbits are mainly situated to right outer part of
the $(R,M_{\rm n})$-plane.  Furthermore, the 2:3, 4:3 and 6:5 resonances
produce thin vertical stability layers.  Furthermore, our numerical
calculations indicate that in the interval $396 \leq M_{\rm n} \leq 407$
there is no indication of the 2:3 resonance.  This justifies why the
stability islands of this resonance were found absent in Fig.
\ref{grd}e when $M_{\rm n} = 400$.  It is also observed, that several
families of higher resonant orbits are present, corresponding to thin
filaments of initial conditions living inside the box region.  We would
like to note, that the maximum value of the $R$ coordinate $(R_{\rm
max})$ is slightly reduced as the nucleus gains mass.  We must also
point out that the $(R,M_{\rm n})$-plane contains only such orbits
starting perpendicularly to the $R$-axis, while all types of orbits
whose initial conditions are pairs of position-velocity (i.e., the 1:1
resonant family) are obviously not included.

\sectionb{5}{CONCLUDING REMARKS}
\label{conc}

In the present work we have sought to shed some light on the interesting
phenomenon of mass transportation by investigating the orbital dynamics
of a mean filed galaxy model, when the mass parameters are linearly
changing in time.  For this purpose, we used an analytic, axially
symmetric, time-dependent galactic gravitational model which embraces
the general features of a disk galaxy with a dense, massive, central
nucleus.  During the galactic evolution the total mass of the galaxy
remains constant which means that whatever mass the disk loses, it  is
gained by the nucleus.  In order to simplify our numerical calculations
we chose to work in the meridional $(R,z)$ plane, thus reducing
three-dimensional to two-dimensional motion.  We kept the values of all
the other parameters constant, because our main objective was to
determine the influence of the mass of the nucleus on the percentages of
the orbits, where mass is transported from the disk to the nucleus.  Our
thorough and detailed numerical analysis suggests that the level of
chaos, as well as the different regular families, are indeed very
dependent on the galactic evolution.  Furthermore, we wanted to prove
that transitions from regularity to chaoticity and vice versa are
possible in this simple model.

Since a distribution function of the model potential was not available
so as to use it for extracting different samples of orbits, we had to
follow an alternative path.  We defined for several time points of the
galactic evolution, dense grids of initial conditions $(R_0, \dot{R_0})$
regularly distributed in the area allowed by the corresponding value of
energy on the phase space.  To show how the mass transportation
influences the orbital structure of the system, we presented for each
case the color-coded grids of initial conditions, which allow us to
visualize what types of orbits occupy specific areas in the phase space.
Each orbit was numerically integrated in the time-independent
Hamiltonian for a time period of $10^4$ time units ($10^{12}$ yr), which
corresponds to a time span of the order of hundreds of orbital periods.
The particular choice of the total integration time was made in order to
eliminate sticky orbits (classifying them correctly as chaotic orbits)
with a stickiness at least of 100 Hubble times.  Then, we made a step
further in an attempt to distribute all regular orbits into different
families.  Therefore, once an orbit has been characterized as regular
applying the SALI method, we then further classified it using a
frequency analysis method.  For the numerical integration of the grids
with the initial conditions of the orbits, we needed about between 5 and
7 days of CPU time on a Pentium Dual-Core 2.2 GHz PC, depending on the
rate of regular orbits in each case.

 The most important outcomes of our numerical investigation can be
summarized as follows:

$\bullet$ Numerous types of ordered orbits were identified in our disk
galaxy model, while there are also extended chaotic domains separating
the areas of regular motion.  In particular, a plethora of resonant
orbits (i.e., 1:1, 2:1, 2:3, 4:3, 6:5 and other resonant orbits) are
present, thereby enriching the orbital structure of the galaxy.  It
should be clarified that by the term ``other resonant orbits" we refer
to resonant orbits with a rational quotient of frequencies made from
integers $> 5$, which of course do not belong to the main families.

$\bullet$  It was observed that the galactic evolution, where mass is
linearly transported from the disk to the central nucleus, influences
mainly the percentages of box and chaotic orbits in the phase
$(R,\dot{R})$ space.  The mass of the central nucleus, although
spherically symmetrical and therefore maintaining the axial symmetry of
the entire galaxy, can generate substantial chaotic phenomena in the
meridional plane, as it is above zero.

$\bullet$  We found that as the galactic nucleus becomes more and more
massive the percentage of chaotic motion grows mainly at the expense of
box orbits, while chaotic orbits is the dominant family once the mass of
the nucleus has reached about 7\% of the mass of the galactic disk.  At
early stages of the mass transportation, where the mass of the nucleus
is still relatively low, we measured the largest amount of ordered
orbits.  In fact, when $M_{\rm n} < 100$ more than half of the phase
space is covered by initial conditions of box orbits.

$\bullet$  Our results strongly indicate that in the time-dependent
Hamiltonian system, where a massive nucleus is developed in the central
region of the disk galaxy through the mass transportation, the character
of the orbits can change either from ordered to chaotic and vice versa
or not change at all.

$\bullet$  The SALI method was proved highly efficient and accurate in the
identification of chaos in the time-dependent system.  Our computations
revealed that this indicator is especially suited for detecting time
intervals where an orbit exhibits a fundamental change in its
character.  Specifically, by following the time-evolution of SALI one
can determine in detail the orbit's successive transitions from
regularity to chaoticity and vice versa.

Judging by the interesting findings we may say that our task has been
successfully completed.  We hope that the present numerical analysis and
the corresponding results will be useful in the field of time-dependent
galactic Hamiltonian systems.  This is a promising step in the task of
understanding the galactic evolution of disk galaxies with spherical
nuclei.  Taking into account that our results are encouraging, we are
planning to properly modify our galactic model in order to expand our
investigation into three dimensions and explore the entire
six-dimensional phase space.  In addition, N-body simulations may
elaborate the mass transportation and its implication on the evolution
of galaxies.


\thanks{ The author would like to express his warmest thanks to the
anonymous referee for the careful reading of the manuscript and for all
the apt suggestions and comments which allowed us to improve both the
quality and the clarity of the paper.}

\References
\vspace*{2\baselineskip}

\begingroup
\renewcommand{\section}[2]{}

\endgroup

\end{document}